\documentclass[letterpaper,twocolumn,10pt]{article}
\usepackage{usenix}

\usepackage{tikz}
\usepackage{amsmath}
\usepackage{xcolor}
\usepackage{graphicx}
\usepackage{subcaption}

\usepackage{filecontents}

\begin{document}

\newcommand{\com}[1]{\textcolor{red}{#1}}
\newcommand{\tool}[0]{Pancake}
\newcommand{\pzf}[1]{{\it \color{blue} pzf: #1}}

\date{}


\title{\Large \bf \tool{}: Hierarchical Memory System for Multi-Agent LLM Serving}


\author{%
Zhengding Hu,
Zaifeng Pan,
Prabhleen Kaur,
Vibha Murthy,
Zhongkai Yu,
Yue Guan, \\
Zhen Wang,
Steven Swanson,
Yufei Ding \\
Computer Science and Engineering, University of California, San Diego
}

\maketitle

\begin{abstract}


In this work, we identify and address the core challenges of agentic memory management in LLM serving, where large-scale storage, frequent updates, and multiple coexisting agents jointly introduce complex and high-cost approximate nearest neighbor (ANN) searching problems. We present \tool{}, a multi-tier agentic memory system that unifies three key techniques: (i) multi-level index caching for single agents, (ii) coordinated index management across multiple agents, and (iii) collaborative GPU–CPU acceleration. \tool{} exposes easy-to-use interface that can be integrated into memory-based agents like Mem-GPT, and is compatible with agentic frameworks such as LangChain and LlamaIndex. Experiments on realistic agent workloads show that \tool{} substantially outperforms existing frameworks, achieving more than 4.29× end-to-end throughput improvement. 


\end{abstract}

\section{Introduction}

Agents have emerged as one of the defining paradigms of the LLM era, enabling complex task scenarios including task planning~\cite{huang2022planner, shinn2023planner2}, knowledge organization~\cite{edge2024graphrag, asai2024selfrag}, tool-augmented generation~\cite{schick2023toolformer, yao2024taubench} and even scientific researches~\cite{wang2023science1, lu2024science2}. These complex tasks, often carried out through multi-turn environment interaction~\cite{yao2022react}, self-reflection~\cite{shinn2023reflexion}, or multi-agent collaboration~\cite{li2023camel}, has introduced substantial information into the generation process and poses significant challenges for context length and attention fidelity~\cite{liu2024lostinthemiddle}. 

In response to this trend, \textbf{Agent Memory}~\cite{zhang2404memorysurvey} has emerged as a key mechanism to manage complex contexts and enhance generation quality. Unlike prior Retrieval-Augmented Generation (RAG) methods~\cite{asai2024selfrag, xu2024recomp, jiang2023flare, lewis2020rag1, jiang2023rag2, lewis2020ragfirst}, which rely on a static knowledge database and lack the ability to capture an agent’s runtime state, results, and other dynamic information, agent memory maintains an external database that records essential information, including external knowledge~\cite{packer2023memgpt, jiang2023rag2}, action history~\cite{xu2025amem, zhong2024memorybank}, user profile~\cite{wang2025deeppersona, ge2024personahub}, and more. The agent must retrieve the most relevant memory items to guide generation throughout LLM inference, while also dynamically inserting newly generated items for future reference.


\begin{figure}[t]
    \centering

    \begin{subfigure}{0.45\textwidth}
        \centering
        \includegraphics[width=\textwidth]{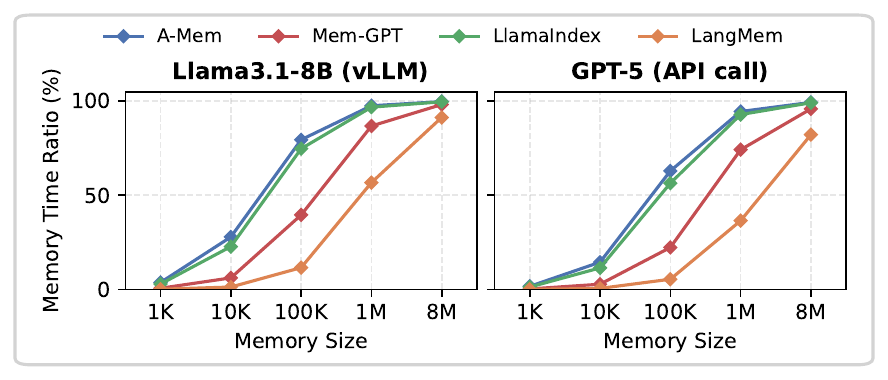}
    \end{subfigure}
    \hfill
    \begin{subfigure}{0.45\textwidth}
        \centering
        \includegraphics[width=\textwidth]{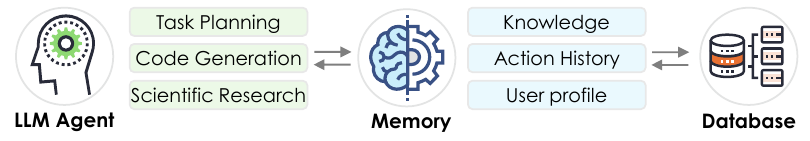}
    \end{subfigure}

    \caption{Memory-based workflow of agentic LLMs.}
    \label{fig:intro-memory}
\end{figure}

\begin{figure}[t]
    \centering
    \includegraphics[width=0.95\linewidth]{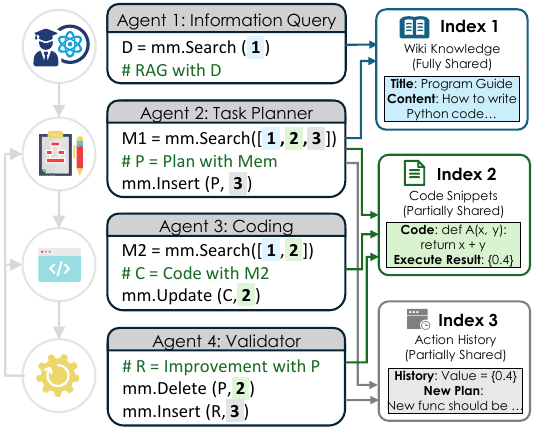}
    \caption{An example of multi-agent memory in \tool{}.}
    \label{fig:interface}
\end{figure}

Agent Memory enables reliable, consistent, and progressively refined agent outputs over complex tasks.
Yet, such a continuous memory mechanism also introduces a highly dynamic database environment that demands frequent approximate nearest-neighbor (ANN) operations~\cite{li2019ann}, typically implemented through embedding vector indexes~\cite{han2023vectordb}. Such operations introduce a new and substantial source of overhead in agent serving when the index scale becomes large. 




Existing agentic memory implementations largely focus on functional support while lacking performance-oriented optimization. As shown in Figure~\ref{fig:intro-memory}, for popular memory-based workflows~\cite{packer2023memgpt, xu2025amem}, the memory operational cost grows sharply with memory size, reaching more than 82\% of the total execution time. Meanwhile, most existing vector database systems fall short in supporting agentic memory: they either optimize only a static index~\cite{douze2025faiss, chen2021spann, jayaram2019diskann, zhang2023vbase, xu2025tribase}, or rely on batch-oriented updates~\cite{mohoney2025quake, mohoney2024incrementalivf, xu2023spfresh, singh2021freshdiskann} designed for periodic maintenance in traditional databases~\cite{vrandevcic2014wikidata}, making them ill-suited for highly dynamic and fine-grained memory operations.


In this work, we present \tool{}, a multi-tier memory management system designed to address the key challenges of implementing an agentic memory system:


First, for a \textbf{single agent}, a key limitation of existing vector search systems is their inability to efficiently handle the frequent, fine-grained updates characteristic of memory workloads~\cite{packer2023memgpt, xu2025amem}.
Existing incremental indexing methods~\cite{xu2023spfresh, mohoney2024incrementalivf} are designed for large-batch insertions in traditional databases~\cite{vrandevcic2014wikidata, kimball2004etl} and rely on direct in-place inserts with periodic rebalancing.
Under the small-batch insertion patterns and interleaved search of agent workloads, inserted vectors are often scattered across various clusters due to high-dimensional distance concentration~\cite{franccois2007concentration} despite strong semantic coherence, degrading both search efficiency and recall.

To address this inefficiency, we explicitly incorporate agent-specific access behaviors into index construction and maintenance.
Specifically, \tool{} exploits both intra-agent and inter-request locality through a multi-level cache index that progressively promotes related vectors to upper levels, improving search ordering and enabling early termination. To guide caching behavior, \tool{} models each agent’s memory access pattern as finite-state machine (FSMs) with continuous updating and merging, enabling index cluster construction closely aligned with the agent’s workload.

Second, supporting \textbf{multi-agent workloads} is challenging, as they frequently invoke different sets of agents to perform memory searches at runtime~\cite{park2023aitown}. 
Using conventional two-level index structures~\cite{douze2025faiss} while maintaining separate indexes for each agent is inefficient. 
At the upper level, this design must traverse the coarse index of every agent, even when only a subset is involved in the query, leading to excessive coarse-search overhead. 
At the lower level, it uses uniform clusters for the shared memory part, ignoring the inconsistent access patterns across agents, thus causing misaligned cluster organization and inefficient fine-grained search.

\tool{} addresses these challenges with a hybrid graph that connects multiple agents’ indexes into a unified structure, enabling the upper-level coarse search to be performed through a single graph traversal. \tool{} also records agent-specific access patterns for each cluster, to reduce cross-agent search overhead caused by inconsistent access patterns.
%
%

For programmability in multi-agent applications, \tool{} provides a simple Python interface that supports operations over arbitrary memory scopes, including different shared and local memory parts, which existing frameworks do not offer. Figure~\ref{fig:interface} shows a multi-agent code-generation setup, where different agents flexibly operate over knowledge, code, and history memories with \tool{} interface.

Third, modern LLM serving systems are typically deployed at \textbf{GPU-CPU platforms}~\cite{kwon2023vllm, zheng2024sglang}, which creates an opportunity to accelerate memory operations. Existing vector databases can only reside entirely on the GPU\cite{johnson2019faissgpu, karthik2025gpuframework2}, or GPU caching mechanism for static indexes~\cite{tian2025gpu-cpu, zhang2024gpuframework1}. However, in agentic serving scenarios, the coexistence of large-scale memory bases~\cite{packer2023memgpt} and LLM inference engine severely restricts available GPU memory. More importantly, the frequent memory updates makes static caching techniques infeasible to apply. To fully utilize resources, \tool{} implements CPU–GPU coordinated index management to accelerate hotspot cluster computation, with an insertion buffer design and asynchronous transfers for low-latency online updates.

In summary, the contribution of this paper is as follows:

\begin{itemize}
    \item We introduce \tool{}, the first multi-tier memory management system tailored for multi-agent applications. \tool{} exploits agent workload characteristics to optimize update strategy for single-agent memory, cluster construction for multi-agent coordination, and dynamic CPU–GPU collaborative execution.

    \item \tool{} can be directly integrated into agent workflows like Mem-GPT~\cite{packer2023memgpt} and plugged into mainstream agentic frameworks like LangChain~\cite{langChain} and LlamaIndex~\cite{Liu_llamaindex_2022}. It provides a concise API through which agents can perform memory operations across flexible memory scopes.

    \item Extensive experiments across diverse agent datasets show that \tool{} delivers over 4.29$\times$ average end-to-end performance speedup compared with existing memory libraries, and reduces the memory-operation time share to an average of 3.2\% under large-scale database.
    
\end{itemize}

\section{Background and Related Work}

\subsection{Memory-based Agent and ANN}





\begin{figure}[t]
    \centering
    \includegraphics[width=\linewidth]{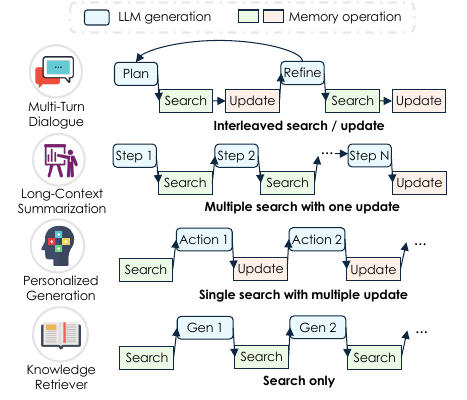}
    \caption{Memory-based agents and their workflows.}
    \label{fig:memory_workflow}
\end{figure}

A memory-based agent typically perform three operations: LLM Generation; Memory Search, which retrieves items relevant to the current context; and Memory Update, which inserts, deletes, or modifies items in the memory store. As shown in Figure~\ref{fig:memory_workflow}, different agent roles induce different operation patterns.
For instance, multi-turn dialogue agents search and update memory at every step to remain consistent with the interaction history~\cite{packer2023memgpt, yu2025memagent}, whereas context-summarization agents search and generate for several rounds and then insert a compressed memory item~\cite{chhikara2025mem0, ouyang2025reasoningbank, zhang2025gmemory}.


These search and update memory operations inherently introduce 
requirements for approximate nearest neighbor (ANN) queries. ANN is typically 
implemented through vector databases, where textual information is encoded into vector embeddings~\cite{wang2022e5,devlin2019bert} and relevance is quantized based on vector 
similarities~\cite{chowdhury2010similarity}. Such ANN queries occur repeatedly throughout an agent’s step-wise generation, their latency and accuracy therefore become increasingly critical to the overall performance of modern LLM serving systems.


To support memory operations, existing memory-based agents such as Mem-GPT~\cite{packer2023memgpt} and A-Mem~\cite{xu2025amem} provide their own memory implementations, and open-sourced agent frameworks like LlamaIndex~\cite{Liu_llamaindex_2022} and LangChain~\cite{langChain} also offer built-in storage interfaces. However, these modules emphasize functionality and rely on suboptimal indexing and searching implementations. As the memory size grows, their query latency increases sharply, reaching more than 99\% of the end-to-end runtime at scale. This trend highlights the need for a scalable and efficient agentic memory framework.




\subsection{Dynamic Vector Database}


Numerous vector-database frameworks~\cite{douze2025faiss, wang2021milvus, chen2021spann, guo2020scaNN, jayaram2019diskann} have explored techniques for efficient search by organizing vectors into structured storage formats, known as indexes. Among them, the Inverted File (IVF) index~\cite{jegou2010ivf} is widely used: it partition vectors into clusters and rank these clusters by the distance between their centroids and the query. Only the top-$nprobe$ clusters are selected for vector-wise search, making $nprobe$ a tunable accuracy–efficiency trade-off~\cite{ray2025metis}. We refer to cluster selection as \textit{coarse search}, and to the search within the selected clusters as \textit{fine search}. Coarse search typically relies on a Flat index or graph-based indexes such as HNSW~\cite{malkov2018HNSW} or Vamana~\cite{jayaram2019diskann} in large-scale settings.


However, existing frameworks are primarily designed for read-only scenarios like RAG, and therefore assume a static vector database with one-shot, full-index construction. Such designs are incompatible with agentic memory workloads, where frequent updates make reconstruction prohibitively expensive. To support online updates, several dynamic vector-database techniques have been proposed~\cite{singh2021freshdiskann, xu2023spfresh, mohoney2025quake, mohoney2024incrementalivf, xiao2024incrementalhnsw}. For example, SPFresh~\cite{xu2023spfresh} avoids global rebuilding through in-place inserts and lightweight local rebalancing, while Quake~\cite{mohoney2025quake} uses a hierarchical cluster structure and adaptively splits clusters based on access frequency. However, these designs typically assume periodic, batch-oriented updates consistent with traditional database workloads~\cite{kimball2013traditionaldb, vrandevcic2014wikidata}. In contrast, agentic memory serving involves highly frequent updates that interleave closely with search operations, leading to degraded efficiency and accuracy in such systems.

\section{Motivation}

\subsection{Inefficient Update Strategy for Single-Agent Memory Access}

In this section, we analyze single-agent memory access patterns and show that existing vector database maintenance algorithms struggle to efficiently handle frequent-update workloads. In typical agent-serving systems, an agent processes many independent requests, each involving multi-step LLM generation and frequent interleaved search–insert operations. Examples include coding agents receiving continuous user tasks~\cite{yang2024sweagent} and scientific agents analyzing large batches of experimental data~\cite{zhang2024honeycomb}.



\begin{figure}[t]
    \centering

    \begin{subfigure}{0.45\textwidth}
        \centering
        \includegraphics[width=\textwidth]{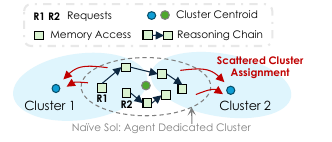}
    \end{subfigure}
    \hfill
    \begin{subfigure}{0.45\textwidth}
        \centering
        \includegraphics[width=\textwidth]{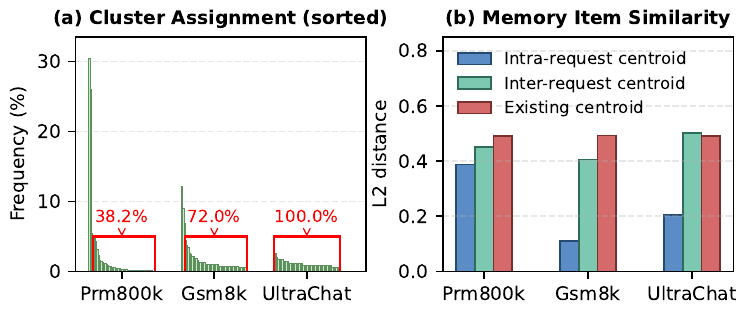}
    \end{subfigure}

    \caption{Direct in-place updates scatter the new vectors into a large number of existing clusters, leading to degradation in efficiency and recall. A naive solution is to leverage intra-agent locality and maintain dedicated clusters for a agent.}
    \label{fig:moti1}
\end{figure}

For vector insertion during memory updates, existing dynamic vector databases~\cite{mohoney2025quake, xu2023spfresh, mohoney2024incrementalivf} typically adopt in-place inserts with periodic updates, as shown in Figure~\ref{fig:moti1}. New vectors are inserted directly into the nearest clusters, with distances calculated between the cluster centroids. Reconstruction is triggered only when the size or the semantic shift~\cite{mohoney2024incrementalivf} of the cluster reaches a threshold. This strategy is effective for large-batch scenarios, while becomes suboptimal with interleaved small-batch search and insert operations.

\noindent \textbf{Scattered Cluster Problem of In-Place Insertion}.
A key issue of in-place insertion is that new vectors inserted into a large pre-clustered index often get scattered across many clusters, even when they are semantically close. As shown in Figure~\ref{fig:moti1}(a), across 100 requests from several agent datasets~\cite{ding2023ultrachat, cui2023ultrafeedback, liu2024apigen}, memory items from the same agent are dispersed into up to 175 clusters, with 38\%–100\% of these clusters being accessed with frequency less than 5\%. This behavior stems from the high-dimensional shell effect~\cite{beyer1999hdtheory1, aggarwal2001hdtheory2}, where points concentrate near the surface of a hypersphere, causing small semantic variations to translate into large differences in centroid distance calculations. Thus, even highly related memory items may be inserted to different clusters.

Such scattered cluster assignments bring challenges for both efficiency and accuracy. First, this forces scanning a larger number of clusters to retrieve semantically related items, incurring extra computation over mostly irrelevant vectors. Second, items in such scattered clusters become harder to locate by centroid distances, and their clusters may be eliminated during the coarse search stage, leading to drop in recall.



To tackle this problem, we first study two locality characteristics of agent memory, which provide critical guidance for designing effective clustering strategies.

\begin{figure}[t]
    \centering

    \begin{subfigure}{0.45\textwidth}
        \centering
        \includegraphics[width=0.95\textwidth]{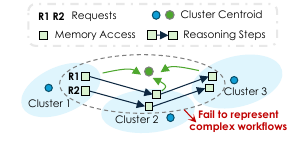}
    \end{subfigure}
    \hfill
    \begin{subfigure}{0.45\textwidth}
        \centering
        \includegraphics[width=\textwidth]{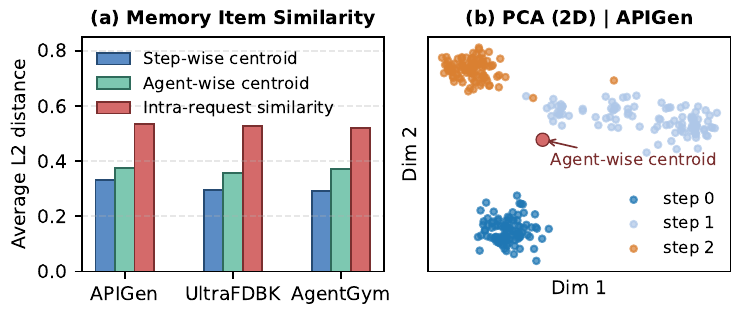}
    \end{subfigure}

    \caption{For more complex workloads, locality across multiple reasoning steps of different requests can be observed. This makes naive dedicated clusters for the agent inefficient, as it fails to capture step-wise clustering.}
    \label{fig:moti2}
\end{figure}

\noindent \textbf{Intra-Agent Locality}. As shown in Figure~\ref{fig:moti1}(b), requests in agentic workflows insert memory items that remain highly coherent across steps and across requests of the same agent. The distances between each item and (i) the centroid of the memory items in the request and (ii) the aggregated centroid of all the agent’s memory items are both substantially smaller than the distances to existing large clusters in the database. This phenomenon is particularly pronounced in task-focused workflows (e.g., mathematical reasoning~\cite{cobbe2021gsm8k}).
%
A straightforward strategy is therefore to assign a dedicated cluster to each agent’s insertion requests. However, this approach is insufficient in more complex workflows, where multi-step reasoning introduces cross-step transitions that cannot be captured by a single cluster (we will detail this in the next paragraph).

\noindent \textbf{Inter-Request Step-wise Locality}. 
Beyond intra-agent locality, more complex workflows reveal an additional layer of structure: memory items from the same reasoning step across different requests tend to cluster together.
For example, in tool-augmented agents~\cite{schick2023toolformer}, the planning, tool-calling, and reflection steps across different requests tend to access similar regions of memory. As shown in Figure~\ref{fig:moti2}(a), memory items belonging to the same reasoning step across different requests demonstrate higher similarity, compared to the intra-request and intra-agent similarity. Figure \ref{fig:moti2}(b) further illustrates the clustering patterns of memory items through 2-dimensional PCA visualization in the tool-calling dataset~\cite{liu2024apigen} with 100 requests, where three clusters emerge, each corresponding to an individual step of the workflow.

This step-wise organization induces frequent transitions across multiple clusters.
Therefore, although maintaining a single dedicated cluster for each agent works for simple workflows in Figure~\ref{fig:moti1}, it is insufficient to capture the step-wise structures in more complicated scenarios, as shown in Figure~\ref{fig:moti2}. 
The green centroid becomes semantically unrepresentative, ultimately degrading accuracy and search efficiency.


In this work, we aim to optimize the agent memory management considering both intra-agent and step-wise locality. Compared to existing approaches~\cite{xu2023spfresh, mohoney2024incrementalivf}, \tool{} introduces more efficient cluster assignment and construction strategies that align with the agent’s complex memory access patterns.

\subsection{Challenges for Multi-Agent Memory Index Management}

\begin{figure}[t]
    \centering
    \includegraphics[width=0.95\linewidth]{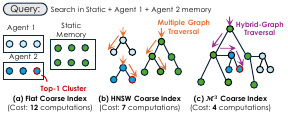}
    \caption{Coarse search costs with different index methods.}
    \label{fig:multiindex}
\end{figure}



Beyond the memory inefficiency for a single agent, we identify the challenges of effectively managing and searching across multiple agent memories, which is a clear need for today’s agentic workloads. In a typical multi-agent setting, each agent continuously updates its local memory, yet may search on the memories of other agents. For example, in generative-agent simulations like AI Town~\cite{park2023aitown}, each agent records its own action and observation histories, but relies on information originating from other agents’ memories to plan behaviors and coordinate group activities. Because different agents become active or interact at different moments, the set of agent memories to search also varies over time.





This brings a clear demand for memory frameworks to support flexible specification of the memory search scope. However, existing ANN libraries~\cite{douze2025faiss, chen2021spann, jayaram2019diskann} only provide interfaces for maintaining and querying on a single index for a given vector database, while offer no native support for search operations across different vector databases.

A straightforward approach is to maintain independent indexes for each agent’s memory. When querying the memories of different scopes, the system searches the corresponding indexes and then merges the results. Although this approach can be implemented directly using existing library interfaces, it suffers from search efficiency issues, described as follows.

\begin{figure}[t]
    \centering
    \includegraphics[width=\linewidth]{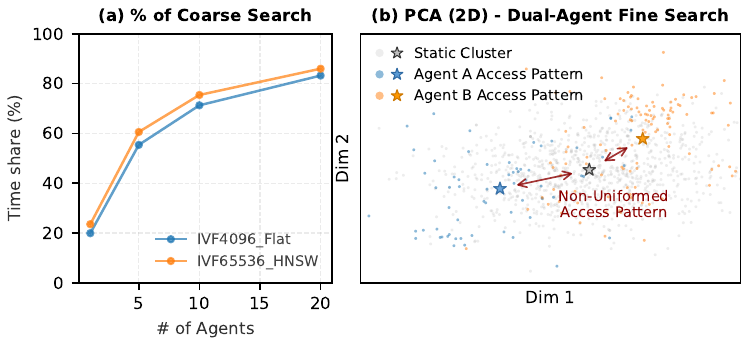}
    \caption{Coarse and Fine Search Challenges in Multi-Agent Memory. (a) Coarse search overhead grows rapidly as the number of agents increases. (b) When two agents access the same cluster in the static memory, their access patterns exhibit non-uniform distributions; circles denote accessed vectors, and stars denote the centroids formed by those vectors.}
    \label{fig:moti3}
\end{figure}







\noindent \textbf{Excessive Coarse Search Cost}. Large-scale vector indexes typically adopt a two-step search: a coarse search first selects the nearest clusters based on centroid distances, and a fine search is then performed within the selected clusters. When independent indexes are maintained for each agent, a wide-scope query must traverse the coarse index of every agent. As shown in Figure~\ref{fig:multiindex}, when querying two agents and the static memory: (a) using a Flat index requires computing the distances to all centroids, and (b) using HNSW~\cite{malkov2018HNSW} requires a full traversal of the coarse-index graph of every agent.

We observe that such multi-index search patterns cause a significant amplification of coarse search cost when the number of agents increases. As shown in Figure~\ref{fig:moti3}(a), with the two commonly recommended Faiss indexes for large-scale settings~\cite{douze2025faiss}, the cost of coarse-grained search rises sharply and exceeds 80\% of the total latency when the number of agents reaches 20. Therefore, it becomes necessary to reorganize coarse indexes across agents, thereby reducing search costs during search across different agent memories.

\noindent \textbf{Non-Uniform Fine Search Patterns across Agents}.
We further observe fine search inefficiency due to the different agent memory access patterns. As shown in Figure~\ref{fig:moti3}(b), when multiple agents query the same cluster in a memory index (constructed with static memory base~\cite{packer2023memgpt}), the vectors they access differ markedly in distribution and clustering behavior. This divergence causes the effective centroid for each agent to shift in the embedding space. This divergence causes the centroids formed by each agent’s accessed vectors within the same cluster to shift noticeably in the embedding space.

This finding indicates that the optimal fine-index organization is highly scope-dependent: for example, an index layout optimized for Agent 1’s access pattern may be poorly aligned with Agent 2’s pattern. As a result, Agent 1’s cluster organization may force Agent 2 to compute over many irrelevant vectors and potentially suffer degraded recall. Such disalignment makes it necessary for coordinated organization of fine indexes across multiple agents.

To address the above issues, \tool{} introduces a hybrid graph for efficient coarse search within only one graph traversal, as illustrated in Figure~\ref{fig:multiindex}. \tool{} also aligns fine-index access by associating each cluster with the pattern recognition of other agents, enabling optimized cross-agent search performance.

\begin{figure}[t]
    \centering
    \includegraphics[width=\linewidth]{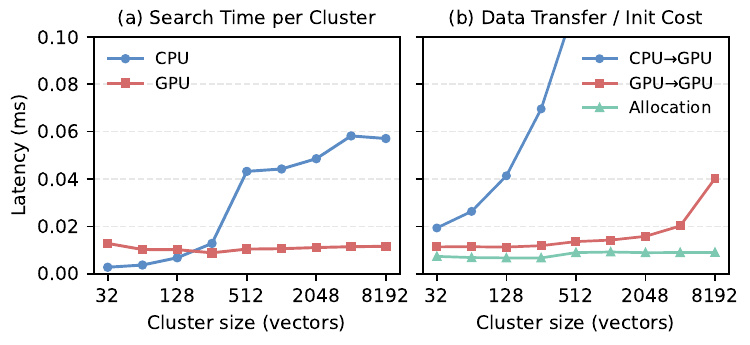}
    \caption{Comparison of operation costs on GPU and CPU, including (a) search time and (b) data transfer and allocation. The results are sampled on the MS MARCO~\cite{nguyen2016msarco} dataset.}
    \label{fig:moti4}
\end{figure}

\subsection{Difficulties for GPU-CPU Collaboration}
\label{sec:moti GPU-CPU}

In this section, we explore how to fully utilize the hardware resources of the GPU-CPU platform, which is widely adopted in LLM inference~\cite{kwon2023vllm, zheng2024sglang}. Vector search involves high-dimensional floating-point computation and therefore benefits substantially from GPU acceleration. As shown in Figure~\ref{fig:moti4}(a), we characterize the performance advantages of CPU and GPU execution. When the number of vectors per cluster is small (< 256), CPU-based computation exhibits lower latency. In contrast, once the cluster size reaches a moderate range ($\ge$ 512), GPU-based vector search achieves a clear speedup of more than $3\times$. The GPU search latency remains largely stable as the cluster grows, since the dominant overhead arises from kernel launch rather than computation. Prior work has extensively explored fully GPU-resident indexes~\cite{johnson2019faissgpu} and search frameworks~\cite{zhang2024gpuframework1}.

However, large-scale vector databases (often over 100 GB~\cite{wikidump}) place heavy demands on GPU memory and make it impractical for the GPU to store the entire index.
The large model weights and KV cache further exacerbate this pressure. This necessitates an on-demand data transfer mechanism between the CPU and GPU. However, such transfer introduces significant overhead, typically far exceeding the actual computation time, as shown in Figure~\ref{fig:moti4}(b). To address this challenge, existing hybrid CPU–GPU designs employ hotspot caching and offloading~\cite{hu2025hedrarag, karthik2025gpuframework2, tian2025gpu-cpu, lin2025telerag} for large-scale indexes. 



The highly dynamic nature of agent memory introduces an additional dimension of complexity for maintaining consistency in CPU–GPU co-managed indexes. Hotspot clusters in agent memory are not only frequently queried but also frequently updated. However, because CUDA lacks efficient mechanisms for concurrent dynamic list expansion, clusters cached on the GPU cannot flexibly support frequent insertions. Prior work primarily supports cache management for static indexes, while performing updates on the CPU index and retransferring the modified clusters back to the GPU incurs prohibitive eviction and transfer costs. 

To address the above challenge, \tool{} implements a GPU–CPU coordinated dynamic index management scheme based on insertion buffers and asynchronous transfers. This design enables dynamically extensible hotspot clusters to be accelerated during both search and update operations.





\section{\tool{}: Methods and System Design}

\subsection{Overview}

In this work, we present \tool{}, a multi-tier ANN-based system designed to meet the demands of dynamic agentic memory workloads. \tool{} follows a coordinated multi-tier design that consists of:
(i) Multi-level, cache-inspired index orchestration informed by trajectory-based agent workload embeddings, enabling locality-aware search and update behavior;
(ii) Multi-layer memory storage that supports efficient sharing, reuse, and migration of memory across agents; and
(iii) Multi-device efficient execution with dynamic hotspot detection and cross-device consistency management to fully leverage heterogeneous CPU–GPU resources.


\subsection{Pattern-Driven Multi-Level Index Cache}

\begin{figure}[t]
    \centering
    \includegraphics[width=\linewidth]{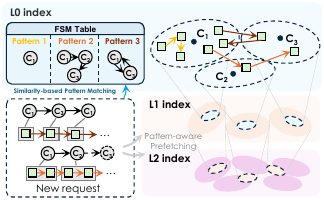}
    \caption{Three-level memory index cache to optimize search efficiency, with FSM-based modeling for access patterns.}
    \label{fig:multi-level}
\end{figure}

\label{sec:multi-level-index}

Existing dynamic ANN methods either rely on streaming insertion and local rebalancing~\cite{mohoney2025quake, mohoney2024incrementalivf, xu2023spfresh}, or on coarse-grained buffering and periodic merging~\cite{zhong2025lsmvec, singh2021freshdiskann}. Both approaches lack awareness of agent-level workload patterns, including intra-request locality and inter-request step-wise locality. 

\noindent \textbf{Three-Level Cluster Caching}. We employ partial caching to resolve the mismatch between localized access operations and the coarse-grained cluster structure of the underlying ANN index. As shown in Figure~\ref{fig:multi-level}, each upper level index forms a subset of the level below, but is organized to more closely reflect the agent’s intrinsic memory-access patterns. 

Search and update over the index always begin at the top level. L0 maintains a table $S$ that tracks the most frequently accessed $N_p$ tiny clusters, which contains the most recently accessed vectors to preserve the agent’s temporal locality. When an L0 cluster overflows, evicted vectors are written back into the L1 index. L1 also maintains $N_p$ intermediate clusters. It caches the top-$k'$ neighbors for each search and update, where $k'$ is slightly larger than the actual retrieval parameter $k$, allowing L1 to store the broader neighborhood around frequently accessed vectors. Finally, once an L1 cluster exceeds a predefined size threshold, it is merged with the L2 clusters, forming a stable and coarse-grained structure.


We leverage the early termination mechanism~\cite{anh2001earlytermination} in vector search to accelerate computation using cached data. During search, whenever all top-$k$ candidates at the current level have distances smaller than $\alpha_{et}\cdot d_{agent}$, we skip computation at the next level. Here, $d_{agent}$ denotes the average top-$k$ distance across recent queries of the same agent. In practice, setting $\alpha_{et}=0.6\sim0.8$ provides a strong trade-off between efficiency and accuracy. We further introduce a verification mode in the system: after an early return, the system optionally performs the complete search in the background. This enables dynamic adjustment of $\alpha_{et}$ without incurring additional latency, while maintaining compatibility with LLM-coordinated speculative-generation workflows~\cite{jin2025ragcache, hu2025hedrarag, zhang2024ralmspec}.

\noindent \textbf{FSM-based Pattern Modeling.}
The agent memory access patterns can be modeled as a Finite-State-Machine (FSM):
\[
P = (S,\, T),
\qquad (c_i \rightarrow c_j) \in T,\; c_i, c_j \in S,
\]
where $S$ is the set of semantic cluster states. Each cluster $(c, \delta) \in S$ stores its cluster centroid $c$ and the average intra-cluster vector deviation $\delta$ from the centroid. The transition set $T$
captures the directed movement of memory accesses across cluster
states. Such FSM abstraction preserves both semantic grouping and step-wise
transition behavior.

The L0 index maintains a pattern table with $N_p$ FSM entries. Given a new
request with memory embedding sequence $(v_1, v_2, \ldots, v_t)$, we compute
its similarity to pattern $P_i$ based on prefix-state alignment and transition
consistency:

\[
\mathrm{sim}(P_i, v_{1:t})
= \sum_{k=1}^{t}
I\big[ (c_{k-1} \rightarrow c_k) \in T_i \big] \cdot
\frac{\delta_k}{1 + |c_k - v_k|},
\]
where $I[\cdot]$ is the indicator function. Using this similarity, each request identifies the best-matching pattern and infers the expected
target cluster for subsequent memory accesses.

\label{sec:multilevel-cache}

\noindent \textbf{Pattern-based Reordering and Prefetching}. Modeling the access pattern enables workload-aware search reordering. For each search operation, the cache manager matches its recent access sequence to a pattern in the FSM table and predicts the most probable L0 and L1 cluster. 
The search process then prioritizes the predicted cluster, enabling a more efficient search order and increasing the likelihood of early termination.

FSM-based modeling also enables prefetch-like behavior in the index cache. After each completed search or update, the system predicts the clusters likely to be accessed next. If these clusters have been evicted or written back, background prefetching is triggered to proactively refresh the cache ahead of time. Prefetching is carried out through an independent search, it can run in parallel with the agent’s LLM-generation steps, creating additional opportunities to reduce overhead.


\noindent \textbf{FSM Construction}. Constructing such FSMs online is challenging, as agent memory accesses arrive in the form of embedding vectors rather than pre-labeled semantic clusters. Classical pattern-recognition approaches (e.g., PCA~\cite{abdi2010pca}, HMMs~\cite{rabiner2002hmm}) are prohibitively expensive for high-dimensional and fine-grained online agent workloads. We therefore adopt a lightweight heuristic FSM construction and merging strategy.

When an agent request completes, the cache system first attempts to match it against an existing FSM in the table. If no match is found, a new FSM is created. During creation, each access in the request sequence becomes an independent state, and states are subsequently merged according to the maximum number of states $N_S$ and the minimum merging distance $d_{\text{merge}}$. If the number of FSM entries exceeds $N_p$, the system merges two FSMs with the highest similarity, producing a compact and continuously updated FSM table. 

\begin{figure}[t]
    \centering
    \includegraphics[width=\linewidth]{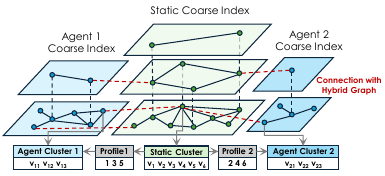}
    \caption{Multi-agent index management with hybrid graph and agent-specific pattern profiling on shared clusters.}
    \label{fig:multi-agent}
\end{figure}

\subsection{Multi-Agent Indexing with Hybrid Graph}
\label{sec:multi-agent}

We propose a multi-agent–friendly index mechanism that incorporates coordinated coarse search and alignment. As shown in Figure~\ref{fig:multi-agent}, our design unifies the multiple coarse indexes into a hybrid graph structure, enabling efficient coarse search through graph traversal. Meanwhile, by associating each cluster with agent-specific memory access patterns, referred to as agent profiles, we further reduce overhead and improve recall for the cross-index operations.

\noindent \textbf{Hybrid Graph Construction}. We introduce a graph structure that connects the static memory and each agent’s local memory. For the fine index, vectors in each static and agent local memory are stored only once, eliminating redundant storage. For the coarse index, each memory scope maintains its own coarse index, organized as a multi-level graph structure similar to HNSW~\cite{malkov2018HNSW}. Each layer forms a bounded-degree graph with up to $M$ neighbors per node. Queries perform a greedy descent through the upper layers, followed by a best-first search at the bottom layer using a frontier of size $ef_{search}$ to approximate the nearest neighbors.

Among multiple coarse indexes, we further introduce \textit{inter-graph connections} to enable navigation across different memory scopes. Specifically, when maintaining each agent’s coarse index, each node in the graph is additionally connected into the static coarse index with probability of $1/ef_{connect}$, thereby creating a controlled number of cross-agent portal nodes that support collaborative multi-agent search.

A cross-scope memory operation begins in the static coarse index entry and performs a BFS-like traversal. When the traversal encounters a node with an inter-connection to the target scope, the search adds the corresponding graph to the search frontier, and set the inter-connected node as the entry. This enables seamless transition across memory scopes while avoiding unnecessary searching over irrelevant regions.



To determine a suitable value for $ef_{\text{connect}}$, we compare the density of the static coarse index with that of each agent-specific index. We measure the average centroid spacing within the private index ($d_{\text{agent}}$) and the static index ($d_{\text{static}}$), and set the inter-connection probability as

\[
ef_{\text{connect}}
    = 
        \min\!\left(
            \alpha_{ic} \cdot \frac{d_{\text{static}}}{d_{\text{agent}}},
            \; 1
        \right).
\]

The intuition is as follows: when the static index covers a broader space, only sparse connections are needed; when the two spaces have similar density, denser connections help avoid cross-graph local minima. Empirically, we choose $\alpha_{ic}$ between 4 and 8 to balance efficiency and recall.


\noindent \textbf{Search Optimization with Agent Profile}. Due to highly non-uniform access patterns, clusters in the static memory exhibit different usage patterns across agents. However, the static memory index cannot adapt to these differences, leading to unnecessary search overhead and preventing the index from aligning with agent-specific access patterns.



To address this, \tool introduces an \textit{agent profile} mechanism for each static cluster. Specifically, every static cluster is associated with an agent-specific table that records the local IDs of recently accessed vectors within that cluster. Such list is maintained as a fixed-size sorted list. Whenever the top-$k$ results of a query fall inside the current cluster, the corresponding vector IDs are promoted to the front of the list. For subsequent accesses, when the agent revisits the cluster, it first retrieves the vectors referenced in its profile by their stored local IDs, enabling a better search order and increasing the likelihood of early termination. Because the profile maintains only vector IDs and a lightweight list structure, the additional storage and management overhead is negligible  comparing to the high-dimensional embedding computations.



\subsection{Dynamic GPU-CPU Index Coordination}
\begin{figure}[t]
    \centering
    \includegraphics[width=\linewidth]{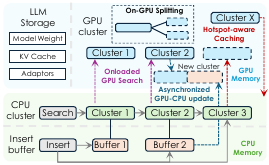}
    \caption{GPU-CPU coordinated index management to enable hotspot cluster computation acceleration.}
    \label{fig:gpu-cpu}
\end{figure}

To further leverage heterogeneous hardware resources, we introduce a GPU–CPU coordinated dynamic index management mechanism, as illustrated in Figure~\ref{fig:gpu-cpu}. Our heterogeneous design consists of a CPU-side insertion buffer and a GPU-side manager for hotspot-aware caching, onloaded search, and consistent cluster maintenance. Such a system enables memory-efficient hotspot acceleration and dynamic cluster organization across devices. This is particularly critical in the co-located serving scenario with LLMs~\cite{hu2025hedrarag}, where the inference engine occupys tens of gigabytes of GPU memory.

\noindent \textbf{Hotspot-aware Caching}. In our hybrid index manager, GPU memory dynamically caches hotspot clusters to accelerate critical computation. For each CPU-resident cluster, the system tracks its access frequency and selects the most frequently accessed clusters according to a predefined GPU memory budget. Whenever the hotspot set changes, the system performs cluster eviction and reallocation to keep the GPU cache aligned with the current workload. Data migration is through asynchronous CPU–GPU transfers to avoid high latency. However, insertions in the agentic memory may cause frequent staleness of the cached clusters.

\noindent \textbf{CPU Insertion Buffer}. As shown in the sampling results of \S~\ref{sec:moti GPU-CPU}, the CPU computation time of a small set of vectors is lower than the GPU’s cluster processing latency. Therefore, we maintain a per-cluster insertion buffer: once a cluster is resident on the GPU, subsequent insertions targeting that cluster are first accumulated in its CPU-side buffer of size $B_{insert}$. For all the searches targeting that cluster, computation is performed collaboratively using both the GPU-cached portion of the cluster and the vectors newly inserted the CPU buffer. The partial results from the two devices are then merged to produce the final results. Because the additional CPU-side search runs in parallel with the GPU computation and contributes only a small fraction of the overall processing time, the end-to-end request latency effectively matches that of a single-GPU execution. Based on this observation, we set $B_{insert}$ to the largest cluster size where CPU-side search cost is lower than GPU-side search, which is 128 on our platform.

\noindent \textbf{Asynchronized Consistency Management}. When the insertion buffer becomes full, the corresponding GPU-cached cluster is resized, and the buffered vectors are migrated from the CPU to the GPU. To avoid the substantial latency caused by on-demand data transfers, we adopt a fully asynchronous cluster-expansion mechanism. The GPU-side index manager proactively allocates new space for clusters to expand and performs data transfers in parallel with online serving. Already cached data are migrated using low-cost GPU–GPU copies, while newly inserted buffer data are transferred through GPU–CPU copies. Once the new data transfer completes, the old GPU cluster is released, enabling seamless online cluster switching. This design eliminates both the waiting overhead associated with synchronous data movement and the memory waste incurred by over-allocating GPU space.

\noindent \textbf{On-GPU Cluster Splitting}. The GPU cache introduces another optimization opportunity: accelerating the computation for cluster splitting. Clustering algorithms like K-means-based methods~\cite{zhao2020kmeans, ding2015yinyang} typically incurs a vector similarity cost that is multiple times higher than that of regular search. Thus, we implement a lightweight kernel based on GPU-based K-means algorithms~\cite{li2013gpukmeans1, bhimani2015gpukmeans2} to onload cluster splitting, avoiding the high computational load and latency on the CPU. Importantly, due to the locality of memory access, clusters that require splitting are usually cached on the GPU, so this technique can reduce the majority of splitting overhead.


\section{Implementation}

\noindent \textbf{User Interface}. \tool{} provides a user-friendly Python interface that exposes simple primitives for agent-memory operations, including search, insert, update, and delete, with explicit specification of the target memory scope. 
Our initialization interface also supports loading from existing indexes, such as Faiss~\cite{douze2025faiss}, enabling reconstruction that is friendly to IVF-based indexes.
Operations submitted through the interface are batched, and adjacent operations of the same type are further grouped into a single batch to improve resource utilization. 

\noindent \textbf{Multi-threaded Index Construction}. In {\tool}, clusters are implemented as multithread-shared data structures, protected by shared-read and exclusive-write locks. Each cluster is associated with metadata, including its index identifier, multi-agent profiles, and its residency status across the multi-level cache and the GPU cache. We maintain a multithreaded execution pool that includes dedicated search threads, update threads, cache-management threads, and GPU-management threads. Insert and delete operations are also handled within the search threads, where items are updated based on the search results. \tool{} adopts asynchronous invocation to ensure concurrency with LLM calls and to maintain compatibility with existing RAG-style systems~\cite{hu2025hedrarag, jin2025ragcache, jiang2024piperag}.


\section{Evaluation}

\begin{figure*}[t]
    \centering
    \includegraphics[width=\linewidth]{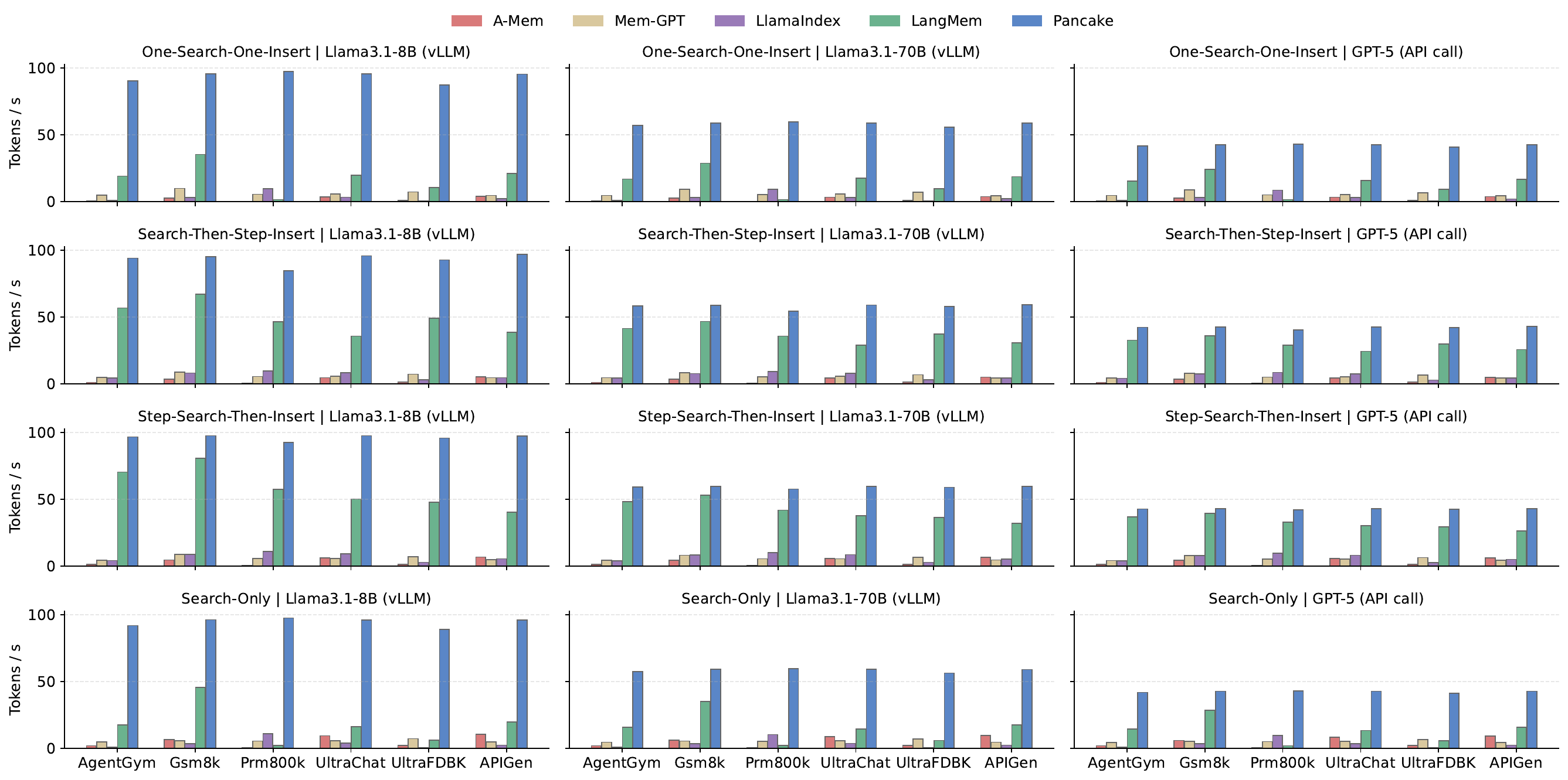}
    \caption{End-to-end throughput comparison between \tool{} and other agentic frameworks, in a single agent scenario across four different access patterns. The experiments are conducted with vLLM~\cite{kwon2023vllm} for Llama models~\cite{touvron2023llama} and API calls for GPT-5. }
    \label{fig:e2e_single}
\end{figure*}

\subsection{Experimental Setup}

\noindent \textbf{Hardware}. We conduct all experiments on a CPU–GPU hybrid server. Each node is equipped with one 64-core AMD EPYC 9534 processor and eight NVIDIA H100 GPUs with 80 GB of memory. The main control, scheduling, and computation logic of {\tool} run on the CPU, while the LLM generation and GPU caching are performed on the H100 GPUs.

\noindent \textbf{Baseline}. For agent serving, we compare four memory design algorithms and their system implementations. These systems provide default ANN-based interfaces for memory management and retrieval. We evaluate them end-to-end by integrating their memory backends with LLM generation workloads. The baselines include:
\textit{A-Mem}~\cite{xu2025amem}: backend for semantically evolving memory in long-term conversational retrieval. \textit{MemGPT}~\cite{packer2023memgpt}: backend for OS-style memory that swaps information between main context and external storage. \textit{LlamaIndex}~\cite{Liu_llamaindex_2022}, vector-store backend in the RAG-oriented framework. \textit{LangMem}: vector-store backend in the agentic framework LangChain~\cite{langChain}.


For evaluating the performance of standalone vector databases, we compare our system against state-of-the-art dynamically updatable vector-index libraries. We use the vectors generated from the memory operations in our end-to-end agent workloads as input to these systems. The baselines include: \textit{Quake}~\cite{mohoney2025quake}, structured and insert-friendly index library with dynamic hot-region–aware optimization. \textit{SpFresh}~\cite{xu2023spfresh}, a large-scale vector search framework based on streaming insertion and localized, balancing-aware reconstruction. \textit{DiskANN}~\cite{jayaram2019diskann, singh2021freshdiskann}, open ANN library that combines an upper-layer graph with a lower-layer cluster index.

For ablation study, we also implement two dynamic maintenance strategies within our framework, including: \textit{\tool{}-IVF-Static}, which initializes an IVF index once and simply appends new vectors to the nearest centroid without any further maintenance. \textit{\tool{}-IVF-Split}, which performs cluster splitting when the size reaches a threshold, consistent with streaming-update and lazy-reconstruction strategies~\cite{mohoney2024incrementalivf}.



\noindent \textbf{Dataset}. We evaluate our system across diverse forms of agent dataset, including multi-turn human–agent dialogue datasets (UltraChat~\cite{ding2023ultrachat}, UltraFeedback~\cite{cui2023ultrafeedback}), long chain-of-thought mathematical reasoning (Prm800k~\cite{lightman2023prm800k}, Gsm8k~\cite{cobbe2021gsm8k}), and task-oriented agent datasets covering function calling (APIGen~\cite{liu2024apigen}) and environment interaction (AgentGym~\cite{xi2024agentgym}).

\noindent \textbf{Workload}. We evaluate several representative memory access patterns, which can be observed in different types of agents:

\noindent --\textit{One-Search-One-Insert}: Each generation step search the memory and updates it with the new output, typical for multi-turn conversational agents~\cite{packer2023memgpt, yu2025memagent}.


\noindent --\textit{Step-Search-Then-Insert}: Each step search the memory, but updates occur only at the end, typical for summarization or long-context compression agents~\cite{chhikara2025mem0, ouyang2025reasoningbank, zhang2025gmemory}.

\noindent --\textit{Search-Then-Step-Insert}: Only the first step performs memory search, while the update occurs in each step, typical for personalized agents driven by user profiles~\cite{ge2024personahub, wang2025deeppersona}.

\noindent --\textit{Search-Only}: The agent only queries memory without updates, typical for RAG-style agents~\cite{asai2024selfrag, xu2024recomp, borgeaud2022retro}.

\noindent \textbf{Static Knowledge Database}. We initialize the vector database similar to the Mem-GPT~\cite{packer2023memgpt} setup, using the MS MARCO corpus~\cite{nguyen2016msarco}, 8M passages in total, as the initial knowledge base. All embeddings are encoded using the E5 model~\cite{wang2022e5} with 1024 dimension.

\subsection{Overall Performance}

In this section, we evaluate the end-to-end improvements on memory-based agents performance when using \tool{}.

\begin{figure}[t]
    \centering
    \includegraphics[width=0.93\linewidth]{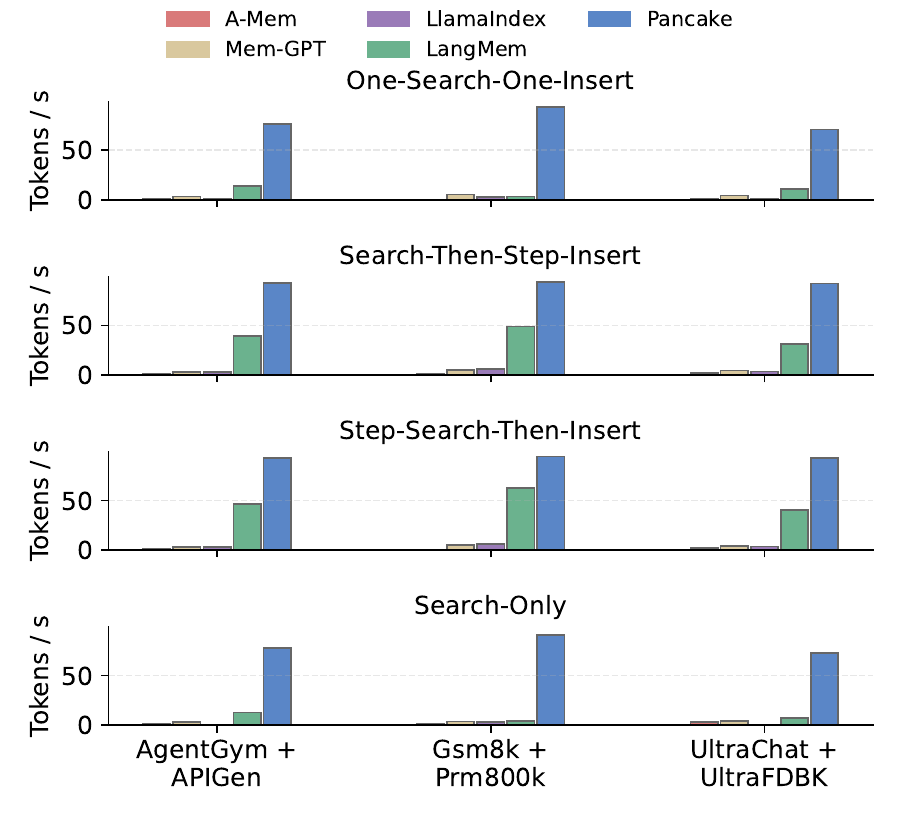}
    \caption{End-to-end throughput comparison in two-agent mixed workload, conducted with Llama3.1-8B.}
    \label{fig:e2e_two_agent}
\end{figure}

\begin{figure}[t]
    \centering
    \includegraphics[width=\linewidth]{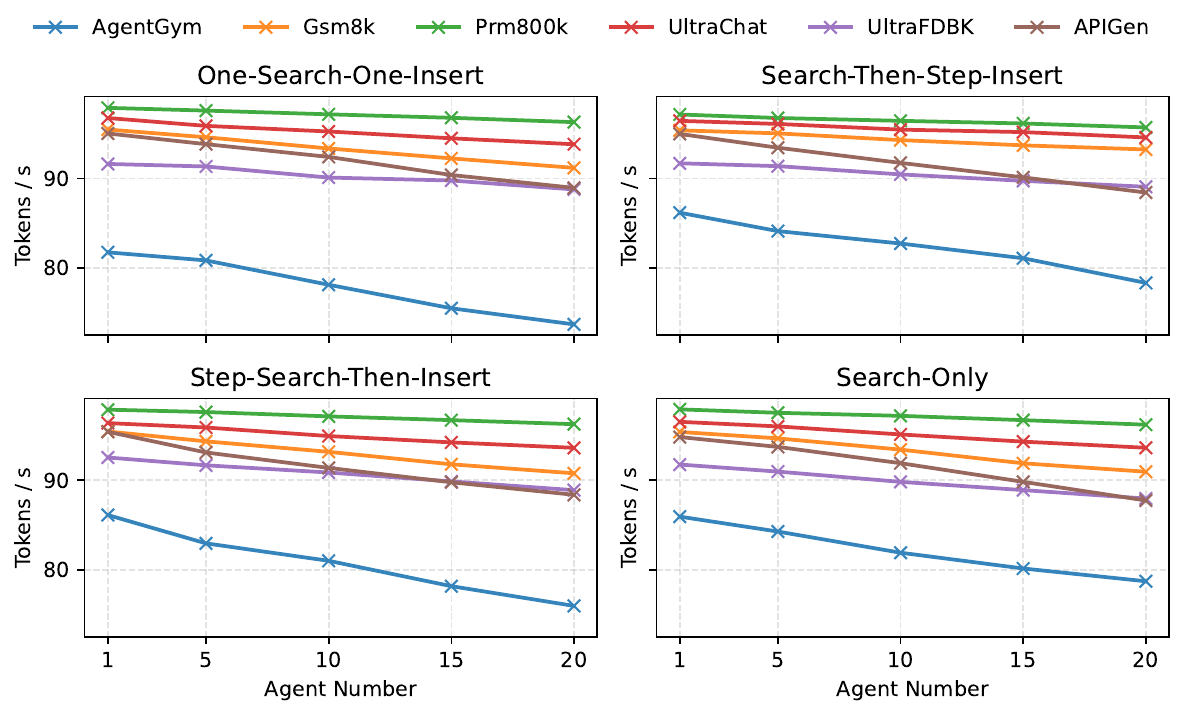}
    \caption{Scaling behavior of end-to-end throughput with an increasing agent number. The experiments are conducted with Llama3.1-8B.}
    \label{fig:agent_scalability}
\end{figure}
\begin{figure}[t]
    \centering
    \includegraphics[width=\linewidth]{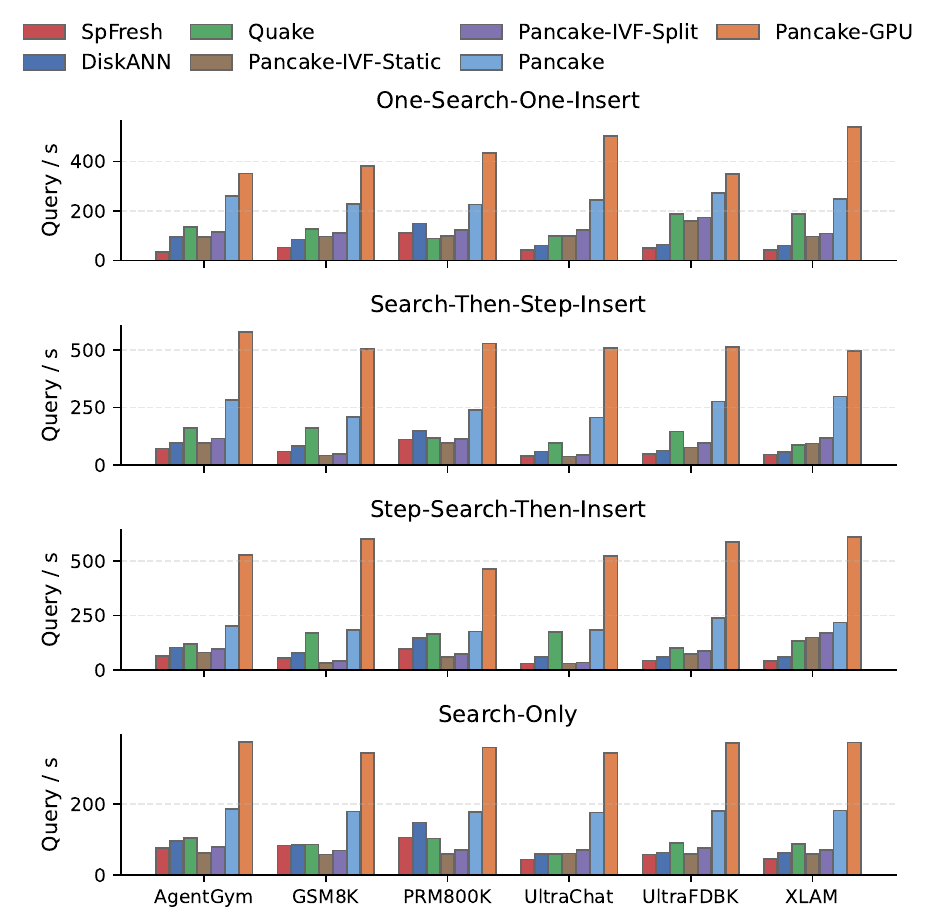}
    \caption{Query throughput of \tool{} and existing vector database implementations, with the batch size set as 8.}
    \label{fig:query throughput}
\end{figure}

\noindent \textbf{Single-Agent Throughput.} We compare different memory management libraries in single-agent settings across multiple models and datasets, as shown in Figure~\ref{fig:e2e_single}.
Across both local inference servers~\cite{kwon2023vllm} and remote API execution, \tool{} consistently sustains stable single-agent request throughput, achieving end-to-end performance improvements ranging from 1.12$\times$ to 26.18$\times$. On average, the speedup over existing libraries is more than 4.29$\times$. 

For the memory operations only, \tool{} achieves speedups of more than 6.81$\times$. The average memory operation time of \tool{} accounts for less than 17.9\%, and on average 3.2\% of the total execution time. This demonstrates the effectiveness of \tool{} in mitigating memory-related bottlenecks.

In addition, we observe that workloads dominated by search operations, including One-Search-One-Insert and Search-only, exert a pronounced performance impact on baseline systems. This is because existing systems rely on suboptimal index constructions and maintenance strategies, which allow low-cost insertions but incur excessive search overhead.

\noindent \textbf{Mixed-Workload Throughput.} We evaluate the impact of two-agent mixed workloads on end-to-end performance, where each agent performs inserts on its private memory and searches on both shared and private memory. As shown in Figure~\ref{fig:e2e_two_agent}, existing memory frameworks exhibit additional performance degradation, dropping by 29.9\%$\sim$55.9\%. This degradation arises from separate memory instance maintenance and the lack of coordinated management across shared and private memory regions, which leads to interference between agents and amplifies operation overhead. In contrast, \tool{} leverages its hybrid-graph design to enable efficient cross-agent search and index alignment, thereby preserving search locality and reducing redundant scans, \tool{} limits the performance drop to no more than 9.8\% under mixed workloads.

\noindent \textbf{Multi-Agent Scalability.} We construct varying numbers of memory-based agents and execute distinct requests over the same dataset, then measure overall throughput with operations across shared and private memory regions. As shown in Figure~\ref{fig:agent_scalability}, \tool{} achieves near-linear scalability in multi-agent settings. With up to 20 concurrent agents (the typical scale of common multi-agent frameworks~\cite{li2023camel, wu2024autogen}), the end-to-end performance degradation remains below 10.2\%.

We also observe that more complex dataset sequences, such as AgentGym and APIGen, exhibit larger performance drops. This is because broader dataset coverage increases the number of nodes traversed during the coarse-level graph search, resulting in proportionally higher search overhead.

\subsection{Comparison with Existing Vector Database}

\begin{figure}[t]
    \centering
    \includegraphics[width=0.98\linewidth]{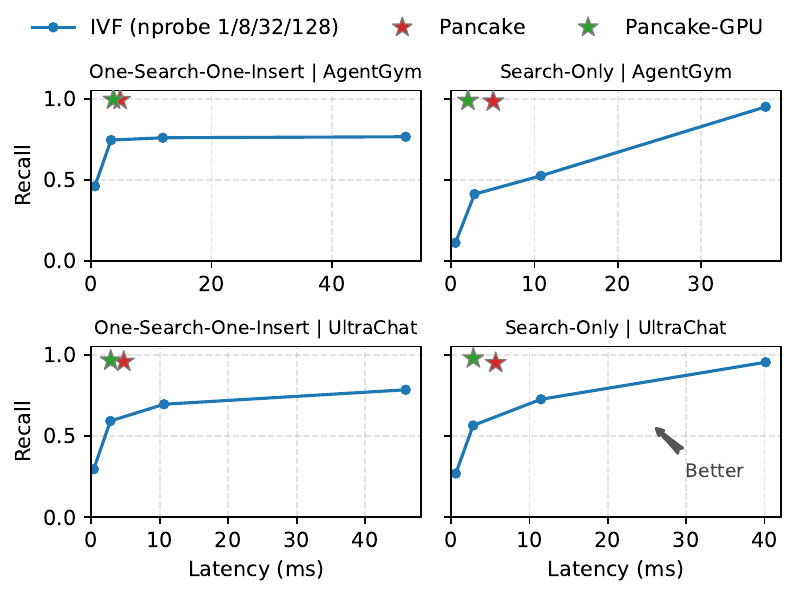}
    \caption{Tradeoff between recall and query latency over different indexing strategies.}
    \label{fig:tradeoff}
\end{figure}

\noindent \textbf{Query Throughput Improvement}. We compare \tool{} with existing vector databases and indexing strategies under online serving workloads. As shown in Figure~\ref{fig:query throughput}, \tool{} consistently improves throughput across memory-intensive agent-serving scenarios, achieving 1.9$\times$ to 4.2$\times$ average speedups over the baselines. These speedups stem from cache and index designs tailored to agent memory-access patterns, which is overlooked in prior work. When leveraging GPU acceleration, \tool{} further achieves an additional 2.2× performance gain, resulting in more than 3.9× speedup over other baselines. This demonstrates the effectiveness of dynamically coordinating GPU resources through our management mechanisms.

\noindent \textbf{Tradeoff between Efficiency and Recall}. We compare recall–latency trade-offs under both mixed search–update and search-only workloads. As shown in Figure~\ref{fig:tradeoff}, directly applying IVF index yields relatively low recall: in the search-only setting, IVF must scan up to 128 clusters to reach recall above 0.9. Under the search–update workload, IVF suffers an even larger recall drop because newly inserted vectors are scattered across different clusters, leading to reduced locality and suboptimal index organization.

By exploiting agent locality and memory-access patterns, \tool{} effectively leverages its caching mechanism to reduce latency while maintaining high recall. Moreover, with coordinated GPU processing, \tool{} can scan additional clusters while simultaneously serving cache hits, achieving lower latency together with a slight improvement in recall.

\subsection{Ablation Study}

In this section, we conduct detailed comparative experiments on the optimization techniques and provide an in-depth analysis of their effects and the root causes of the improvements.

\begin{figure}[t]
    \centering
    \includegraphics[width=0.98\linewidth]{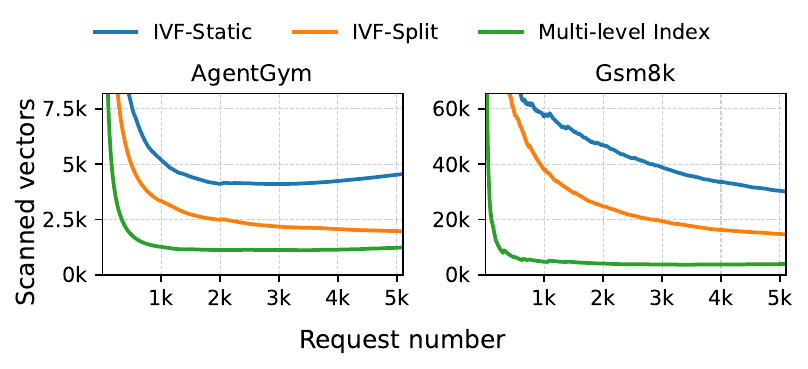}
    \caption{Number of scanned vectors to achieve fully recall of top-5 memory items. The insertion-to-search ratio is 1:1.}
    \label{fig:scanned_stop}
\end{figure}

\noindent \textbf{Optimized Index with Multi-level Cache}. We compare how different dynamic maintenance strategies affect the search costs with dynamism. A lower number of scanned vectors indicates that the index has evolved into a structure better aligned with the current agent’s access pattern and provides stronger early-termination opportunities, thereby improving performance. As shown in Figure~\ref{fig:scanned_stop}, performing IVF-Static updates leads to significantly higher scan counts, because newly inserted memory items are distributed across many clusters rather than being localized. IVF-Split eventually reduces the number of scanned vectors with sufficient insertions and the stable clusters formed. However, the long pre-convergence phase can be observed since the index cannot rebalance until the splitting threshold is reached. In contrast, our multi-level index cache effiently exploits agent-specific spatial and temporal locality, allowing it to stabilize at a low scan cost much earlier. This early adaptation leads to up to 2.23$\times$ latency reduction over long-serving workloads.

\begin{figure}[t]
    \centering
    \includegraphics[width=0.98\linewidth]{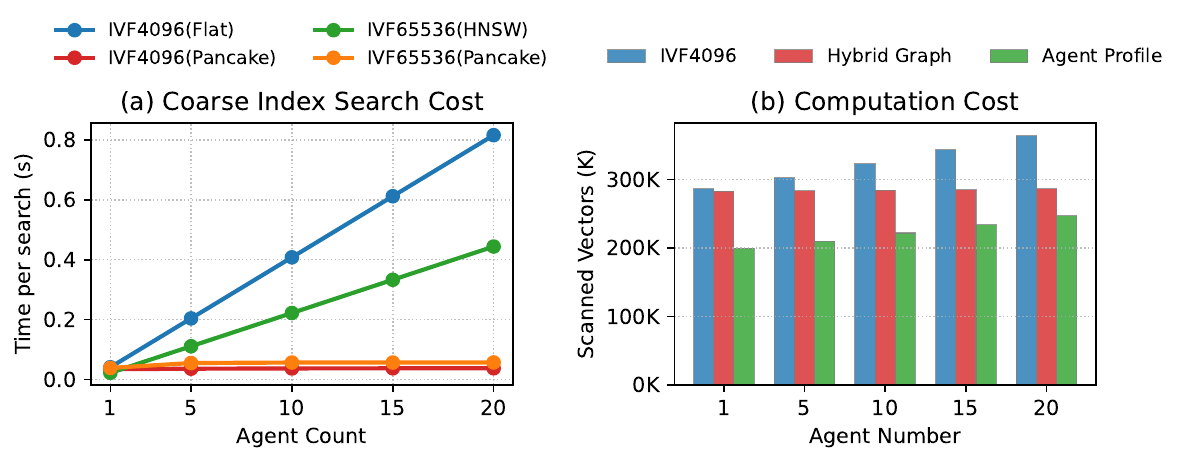}
    \caption{Efficiency improvements from multi-level index management on (a) coarse index search cost and (b) total computation cost.}
    \label{fig:multiagent_ablation}
\end{figure}

\noindent \textbf{Search Efficiency with Multi-Agent Index}. We first compare the reduction in coarse index search cost under multi-agent index management. As shown in Figure~\ref{fig:multiagent_ablation}(a), maintaining separate indexes for each agent leads to a near-linear increase in coarse search overhead as the number of agents grows. In contrast, our multi-index management employs a hybrid graph that interconnects agents’ coarse indexes, enabling efficient navigation of the global search space and achieving more than a 20× reduction in coarse search cost.

We further compare the total search cost under different optimization strategies. As shown in Figure~\ref{fig:multiagent_ablation}(b), the hybrid-graph construction reduces the number of vector similarity computations by up to 11.6\% compared to independently constructed indexes. Moreover, when incorporating agent profiles, we can track each agent’s access preferences within the static clusters, achieving an additional 21.8\% reduction in average computation cost without modifying the global index layout. These results highlight the unique advantages of \tool{} in multi-agent index management.

\begin{figure}[t]
    \centering
    \includegraphics[width=0.98\linewidth]{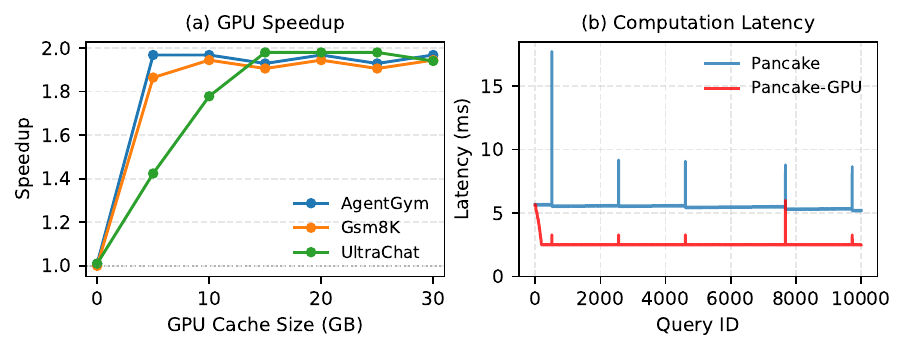}
    \caption{(a) GPU speedups under varying pre-allocated GPU cache sizes. (b) Computation latency of each query over the input workload with AgentGym dataset and 10GB GPU memory cache size. The insertion-to-search ratio is 1:1.}
    \label{fig:gpu_speedup}
\end{figure}

\noindent \textbf{Speedups with GPU Caching}. We evaluate how GPU cache size affects performance. As shown in Figure~\ref{fig:gpu_speedup}(a), the GPU-accelerated version achieves up to 1.92$\times$ speedup over the CPU baseline and reaches a performance plateau with only 5$\sim$15 GB of GPU memory. The effectiveness of GPU acceleration depends on the workload: conversational datasets distribute query-relevant clusters more widely, requiring a larger GPU cache to fully exploit acceleration.

We also examine latency over time under a mixed search–insert workload. As shown in Figure~\ref{fig:gpu_speedup}(b), the GPU version warms up quickly by caching hot clusters and maintains low, stable latency. Occasional spikes arise when streaming insertions trigger cluster splits, temporarily introducing additional computation. In our GPU-enabled design, most cluster operations are performed on the GPU, reducing the cost of these split events and keeping their impact minimal.








\section{Conclusion}

We presented \tool{}, a multi-tier memory management system that bridges the gap between dynamic agentic memory and ANN-based vector indexing. \tool{} leverages semantic locality for single-agent workloads, hybrid indexing for multi-agent memory management, and CPU–GPU collaborative indexing for acceleration. With a simple Python interface and support for flexible multi-scope memory operations, \tool{} integrates easily into existing agent frameworks. Experiments across diverse agent datasets show that \tool{} significantly reduces memory operation overhead and delivers more than 4.29$\times$ average end-to-end speedup over existing implementations.

\bibliographystyle{plain}
\bibliography{ref}

\end{document}